\documentclass[a4paper,11pt, twocolumn]{article}
\usepackage{graphicx}
\usepackage{amsfonts}
\usepackage{amsmath}
\usepackage{listings}
\usepackage{mathtools}

\usepackage{hyperref}

\author{Dunn E. J.$^{1,2}$, Young R. J.$^{1}$ and Jarvis S. P.$^{1,*}$}
\date{%
   \footnotesize $^1$School of Physics and Astronomy, Lancaster University, Lancaster, LA1 4YB, UK.\\%
   $^2$Mathematical and Physical Sciences, University College London, Gower Street, London, WC1E 6BT, UK.\\
   $^*$Corresponding author: \href{mailto:samuel.jarvis@lancaster.ac.uk}{samuel.jarvis@lancaster.ac.uk}
}
\normalsize
\usepackage[font=small,labelfont=bf]{caption}
\usepackage{xr}

\graphicspath{{Figures/}}

\begin{document}
\title{Force-Isosurface Simulations Probe the Limits of High-Resolution AFM on Three-Dimensional Molecules}

\twocolumn[
  \begin{@twocolumnfalse}
    \maketitle
    \begin{abstract}
      \noindent High-resolution atomic force microscopy has transformed molecular imaging by revealing intramolecular structure directly in real space. A major remaining challenge is to extend this capability from largely planar molecules to non-planar molecular systems, where the most important structural information may be distributed across different heights above the surface. Here we use probe-particle-model simulations to predict the constant-force contours expected above molecules with increasing structural complexity. By extracting force isosurfaces from simulated three-dimensional force fields, we compare the molecular information retained in constant-height and constant-force images. For tilted benzene and pyrrole, constant-force images preserve the molecular framework across a range of adsorption angles and allow the molecular orientation to be recovered quantitatively. For larger non-planar and three-dimensional systems, simulations identify characteristic force-isosurface contrast associated with adsorption geometry, lower-lying molecular structure and curved molecular surfaces. These results provide target contrasts for force isosurfaces that could be extracted from three-dimensional force-mapping experiments, evaluating the molecular information retained by ideal force-isosurface imaging across progressively non-planar systems.
      \end{abstract}
  \end{@twocolumnfalse}
]

\paragraph{Introduction.}

High-resolution scanning probe microscopy has provided an unparalleled route to studying molecules assembled on surfaces through direct real-space imaging \cite{JelinikReview, GrossReview}. Such measurements have been used to investigate molecular adsorption, on-surface reactions, heterogeneous catalysis, molecular electronics and the relationship between molecular structure and function \cite{LackingerCOF,HuangSwitch,bondRev}. In particular, non-contact atomic force microscopy (NC-AFM) with functionalised tips has enabled intramolecular contrast with sufficient resolution to assign molecular structure, distinguish bond order and identify subtle differences in adsorption geometry \cite{GrossReview,probeParticle,c60GrossBondOrder,pentaceneDist}. However, the interpretation of high-resolution AFM contrast is rarely straightforward. The measured image is shaped not only by the molecular structure, but also by the tip apex, the force field above the molecule, probe relaxation and the chosen imaging mode. As a result, simulations are essential for connecting observed contrast to molecular structure \cite{probeParticle}.

%
Most submolecular AFM imaging has been performed using constant-height measurements, particularly in ultra-high vacuum (UHV) and often at low temperature, where exceptional mechanical stability and well-defined functionalised tips can be achieved \cite{GrossReview,JarvisIJMSReview}. Constant-height imaging is highly effective for planar molecules, where relevant structural features lie at similar height above the surface. For non-planar molecules, molecular conformers and three-dimensional adsorbates, the situation is more difficult. If the tip height is chosen to resolve the highest part of the molecule, lower-lying regions may contribute weakly or disappear from the image. If the tip is brought close enough to probe those lower regions, the highest parts of the molecule may become unstable or inaccessible. This creates a general challenge: the molecular systems for which three-dimensional structural information is most valuable are often those for which simple constant-height imaging is least well suited.

%
Several important developments have shown that AFM can, nevertheless, provide quantitative information about three-dimensional molecular structure. Schuler \textit{et al.} demonstrated that adsorption geometry can be determined from high-resolution AFM measurements, including molecular tilts and adsorption height differences \cite{schuler2013adsorption}. Subsequent work showed that non-planar molecules and three-dimensional surface objects can be imaged with submolecular resolution by following or reconstructing the molecular force field \cite{OscarCustance}. Non-planar porphyrin conformers have also been directly distinguished by NC-AFM, demonstrating the sensitivity of the technique to subtle conformational differences \cite{JaschaRepp,Jarvis2HTPP}. More recently, three-dimensional intramolecular force mapping has been extended to room-temperature measurements, further demonstrating the structural information available in molecular force fields \cite{BrownRTForceMapping,BrownRTAdsorptionGeometry}. These studies establish that AFM can access three-dimensional molecular information, but they also highlight the experimental complexity of acquiring and interpreting full force maps. Simulations that can accurately predict the expected images, and potential limitations in resolution, are therefore a useful aid when deciding on the most appropriate experimental approach. 


An alternative route is to consider images formed under force-feedback conditions. In an ideal constant-force image, the recorded height does not correspond to a fixed geometric plane above the sample, but to a force isosurface in the tip–sample interaction field. A force isosurface is not a simple topographic map of the molecular skeleton, but instead reflects the spatial distribution of the interaction force above the molecule. In conventional frequency-modulation NC-AFM, this quantity is not measured directly, since the primary observable is the frequency shift of the oscillating cantilever. However, PeakForce-style and related off-resonance methods commonly used in ambient environments acquire force curves during imaging and use force-based feedback to regulate the tip–sample interaction \cite{SuPeakForce,proteinsOffResonance}. These approaches therefore provide an additional experimental motivation for considering the molecular contrast expected from force-isosurface imaging.

\begin{figure*}[h!]
\centering
\includegraphics[width=\textwidth]{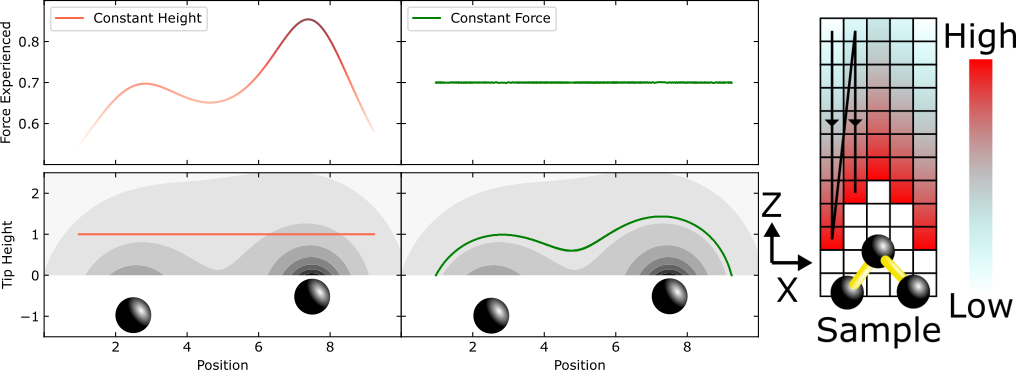}
\caption{A schematic comparison of constant-height and constant-force imaging in an arbitrary force field. The upper panels show the force experienced by a tip following constant-height (\textit{top left}) and constant-force (\textit{top centre}) trajectories, with the corresponding paths shown on force contour maps below. The right panel illustrates the extraction of a constant-force image from a three-dimensional force field. At each $x$-$y$ position, the algorithm follows a vertical column towards the sample and records the highest $z$ value at which the selected force threshold is exceeded. Higher regions of the sample reach this threshold at larger $z$ values, producing a force isosurface that follows the molecular structure from above. \label{cartoonCombo}}
\end{figure*}

Here we simulate constant-force AFM images by extracting force isosurfaces from probe-particle-model force fields. Rather than calculating only conventional constant-height images, we determine the height at which a selected force threshold is reached above each lateral position, producing the image expected from an ideal constant-force measurement. This allows us to compare how molecular information is represented when the tip follows the interaction force field rather than scanning in a fixed plane.

We show that constant-force images can retain submolecular and three-dimensional structural information that is lost in constant-height imaging of non-planar molecules. In simple tilted molecular units, the force isosurface preserves the full molecular geometry and enables quantitative recovery of molecular orientation. In larger molecules, the same approach distinguishes adsorption geometries, reveals lower-lying structural features and resolves contrast around strongly curved molecular surfaces. These results therefore evaluate the molecular information retained by ideal force-isosurface imaging across progressively non-planar systems and provide a route for predicting target contrast associated with three-dimensional molecular structure.

%


\paragraph{Simulating constant-force images.}
Constant-force images were simulated using the Probe Particle Model \cite{probeParticle}. In this model, a flexible probe particle represents the functionalised tip apex and interacts with the molecular structure through Lennard--Jones potentials. Relaxation of the probe particle within this potential gives the force acting on the tip, allowing high-resolution AFM contrast from functionalised tips and nano-asperities to be simulated \cite{FePC-CO,jarvisc60}. A CO-like probe particle and a lateral stiffness of $4$~N\,m$^{-1}$ were used throughout. Electrostatic interactions were omitted, so that the comparison was based on the short-range force contrast arising from the molecular structures.

For each molecule, the tip--sample force was calculated on a three-dimensional grid above the molecular coordinates. At each lateral position, the probe particle was relaxed and the $z$ component of the force acting on the tip was recorded. Repeating this calculation over a series of closely spaced height planes produced a three-dimensional force field above the molecule. Positive force corresponds to repulsive interaction. The vertical grid spacing was $0.01$~\AA, while the lateral spacing ranged from $0.05$ to $0.2$~\AA\ depending on the dimensions of the molecular system.

Constant-force images were extracted from this force field using a custom Python routine. Each vertical force column was searched from the largest tip--sample separation towards the molecule, and the highest sampled $z$ position at which the selected force threshold was exceeded was recorded. This produces the outermost force isosurface corresponding to the contour followed in an ideal constant-force measurement. The procedure is analogous to extracting force information from three-dimensional force-mapping experiments, where frequency-shift spectra are acquired over a grid of lateral positions and converted into force data \cite{pawlak2012high,schuler2013adsorption,OscarCustance,JaschaRepp}. The extraction process is illustrated schematically in Figure \ref{cartoonCombo}.

\begin{figure*}[h!]
\centering
\includegraphics[width=\textwidth]{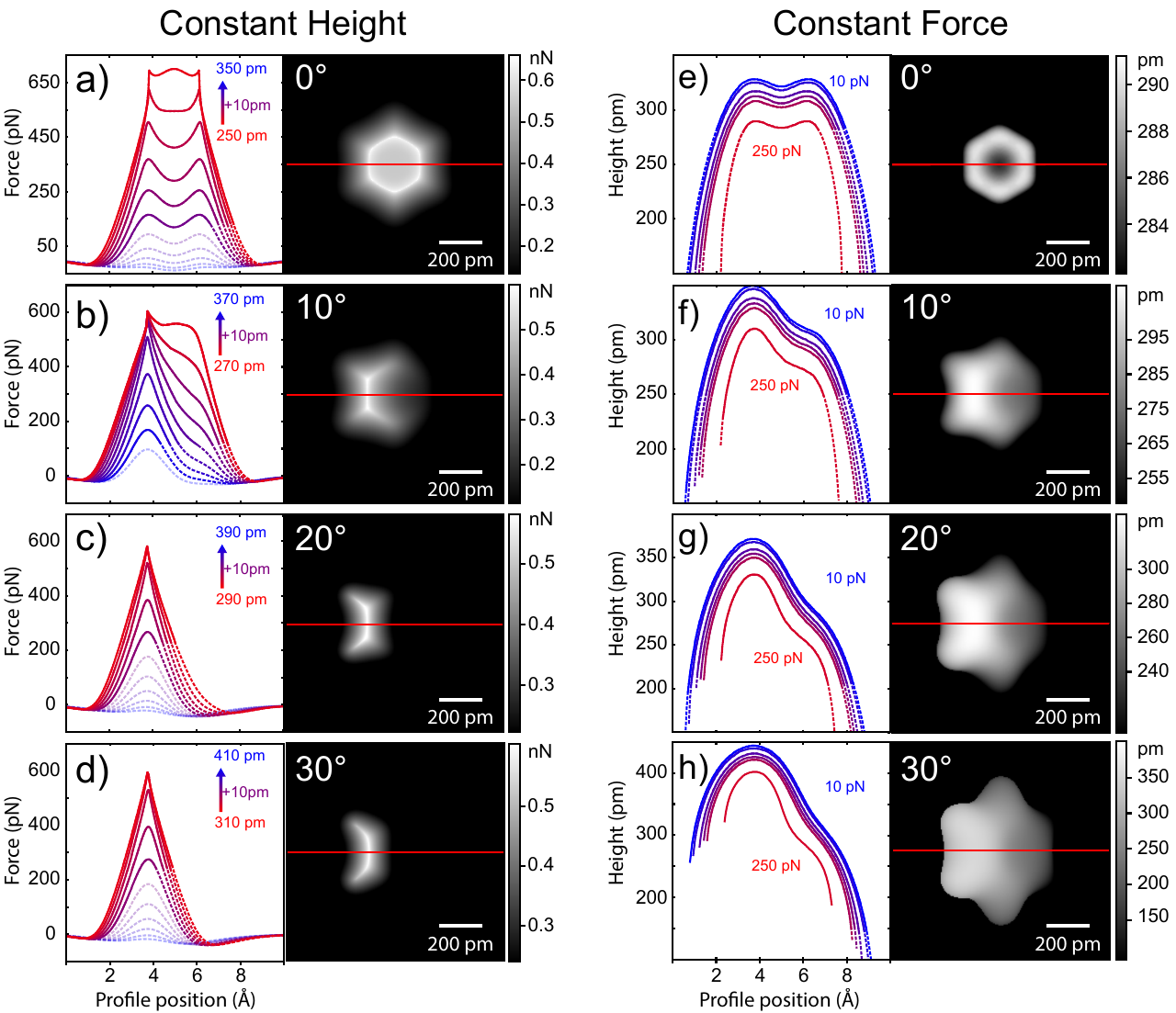}
\caption{Simulated constant-height and constant-force images of benzene tilted by $0^\circ$, $10^\circ$, $20^\circ$ and $30^\circ$, shown in rows 1--4, respectively. Line profiles show the dependence of the apparent features on the selected height or force threshold. Constant-height profiles were calculated at height increments of $0.1$~\AA, while constant-force profiles were calculated at $10$, $20$, $50$, $75$, $100$ and $250$~pN. Pale regions of the profiles indicate values outside the colour scale used for the corresponding image. A series of constant height and constant force images are shown at fixed values $2.6$~\AA\ and $250$~pN, respectively, to provide a visual comparison for molecular tilt. For flat-lying benzene, the constant-height image recorded $2.6$~\AA\ above the molecule resolves a sharp hexagonal ring. As the molecule is tilted, contrast from the lower side of the ring is progressively lost until only the highest atoms remain clearly visible. In contrast, the constant-force images recorded at $250$~pN retain the full ring across the same range of tilt angles.  \label{4x4}}
\end{figure*}

\paragraph{Rotation of benzene and pyrrole.}

Benzene and pyrrole were used as simple model systems for phenyl and pyrrole-like units in larger molecules. To compare constant-height and constant-force imaging, we simulated images of tilted benzene at a range of heights and constant force values as shown in Figure \ref{4x4}. 

The dependence of the simulated contrast on force threshold and lateral probe-particle stiffness is examined in Figure S1; the selected values of 250~pN and 4~N\,m$^{-1}$ lie within a stable parameter range while retaining high-resolution molecular contrast.

The constant-height image of benzene lying parallel to the surface shows a sharp hexagonal ring, with weaker diffuse contrast from the hydrogen atoms at the vertices. The corresponding constant-force image appears broader, reflecting the contour followed by the tip as it maintains the selected force threshold. As the molecule is tilted, the difference between the two approaches becomes more pronounced. In constant-height images, contrast from the lower side of the ring rapidly decreases, and at a tilt angle of $30^\circ$ only the highest carbon atoms remain clearly visible. In constant-force images, the full ring remains visible across the same range of tilt angles because the tip follows the molecular force field rather than remaining in a fixed plane. The resulting force isosurface therefore retains more of the molecular geometry for tilted structures. The profiles in Figure \ref{4x4} further show that the absolute contour height and lateral width vary with the selected force threshold, while the complete tilted ring remains represented over the range $10–250$ pN.

\begin{figure}[h!]
\centering
\includegraphics[width=0.5\textwidth]{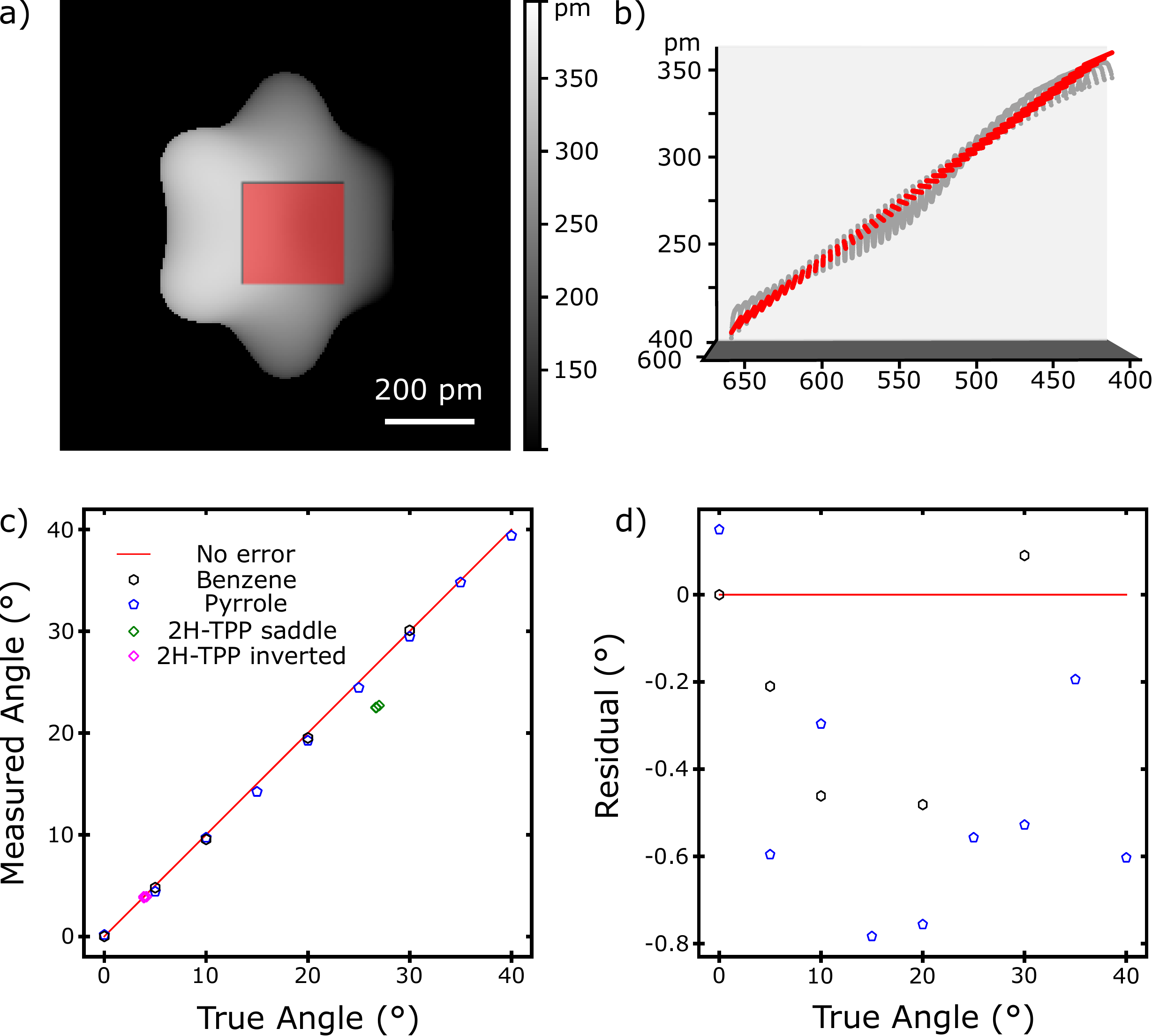}
\caption{Plane fitting of simulated force isosurfaces for benzene, pyrrole and 2H-TPP. A plane was fitted to a selected region of each force isosurface to determine the molecular tilt angle. Panels \textit{(a)} and \textit{(b)} show the fit for benzene tilted by $30^\circ$, with the grey points representing the simulated force isosurface and the red points representing the fitted plane. The measured tilt angles for benzene (\textit{black hexagons}) and pyrrole (\textit{blue pentagons}) closely follow the true input angles \textit{(c)}, with errors below $1^\circ$ across the full range studied \textit{(d)}. The inverted geometry of 2H-TPP (\textit{green diamonds}) follows the same trend, whereas the saddle geometry (\textit{magenta diamonds}) gives a systematically lower measured angle. \label{planeFitResults}}
\end{figure}

Schuler \textit{et al.} showed that adsorption angles can be determined by fitting a plane to frequency-shift isosurfaces over small planar molecules such as olympicene \cite{schuler2013adsorption}. We adopted an analogous approach here by tilting the atomic coordinates of benzene and pyrrole and fitting planes to the resulting force isosurfaces. As shown in Figure \ref{planeFitResults}, the true tilt angle was recovered to within $1^\circ$ for both molecules over the full range studied.

These results illustrate the trade-off between constant-height and constant-force imaging for tilted molecular structures. Constant-height imaging provides sharper contrast for the uppermost atoms, but the image becomes increasingly incomplete as the molecular tilt increases. Under the conditions simulated here, the lower side of benzene is largely lost by $10^\circ$. Constant-force imaging gives broader contrast, but more accurately recovers the molecular structure and allows the tilt angle to be recovered quantitatively over the range studied. For simple molecular units, the force isosurface therefore provides a more reliable representation of molecular orientation than a single constant-height image.

\paragraph{Differentiating between adsorption geometries of 2H-TPP.}

Having established the approach for simple tilted molecules, we next applied the same analysis to two adsorption geometries of 2H-tetraphenylporphyrin (2H-TPP). 2H-TPP consists of a porphyrin macrocycle with four pyrrole-like units and four peripheral phenyl groups. The nitrogen atoms point towards the centre of the macrocycle, where two hydrogen atoms occupy two of the four nitrogen sites \cite{porphyrinReview}.

\begin{figure}[h!]
\centering
\includegraphics[width=0.5\textwidth]{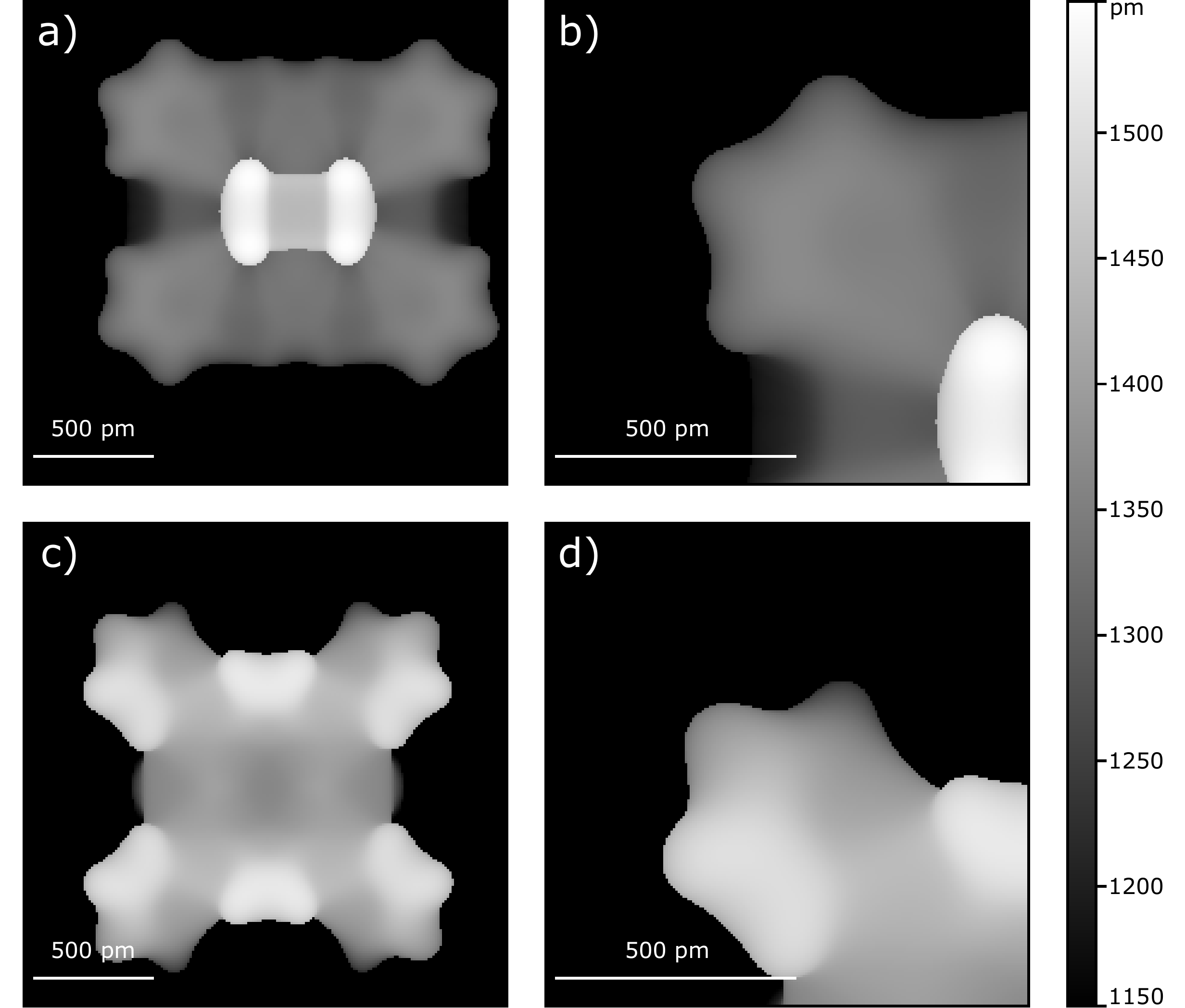}
\caption{Constant-force images of 2H-TPP in the `inverted' \textit{(a)} and `saddle' \textit{(c)} adsorption geometries. In the inverted geometry, the pyrrole-like units at the centre of the molecule protrude strongly from an otherwise near-planar structure. Plane fitting to the phenyl-ring region indicated in \textit{(b)} gives an angle of approximately $4^\circ$. In the saddle geometry, the same analysis gives a phenyl-ring angle of approximately $22.5^\circ$ \textit{(d)}. \label{2HTPP}}
\end{figure}

Figure \ref{2HTPP} shows force isosurfaces of 2H-TPP simulated at a $250$~pN setpoint using coordinates from DFT models of two experimentally observed geometries \cite{Jarvis2HTPP}. In the `saddle' geometry, the pyrrole-like rings alternate between upward and downward orientations. Two of the pyrrole rings therefore protrude above the macrocycle and appear as lobes in the force isosurface. The phenyl rings are rotated towards these raised pyrrole units, producing six raised features arranged in two groups of three. The porphyrin centre, occupied only by two hydrogen atoms, appears lower in the image as a central void.

By contrast, in the `inverted' geometry, two pyrrole-like rings protrude almost vertically from the molecule and dominate NC-AFM images of the structure \cite{JaschaRepp, Jarvis2HTPP}. In the simulated constant-force image, the phenyl rings and the remaining pyrrole-like rings, which lie close to parallel with the surface, are also visible. The centre of the porphyrin is less clearly resolved because of the large force contribution from the protruding pyrrole rings. This strongly distorted geometry gives the characteristic rectangular appearance observed previously in STM and NC-AFM \cite{JaschaRepp, Jarvis2HTPP}.

For the `inverted' geometry, plane fitting gives phenyl-ring angles of $3.9$, $3.9$, $3.8$ and $3.9^\circ$ (error $\pm 0.5^\circ$), in good agreement with the values calculated directly from the coordinates: $4.1$, $4.1$, $3.9$ and $3.9^\circ$, respectively.

For the `saddle' geometry, the same analysis gives phenyl-ring angles of $22.5$, $22.5$, $22.7$ and $22.5^\circ$ (error $\pm 0.9^\circ$), compared with true values of $26.7$, $26.7$, $27.0$ and $26.7^\circ$. The fitted angles are therefore underestimated by approximately $5^\circ$. Since the tilt of isolated benzene is recovered accurately, this discrepancy is attributed to the influence of the surrounding molecular structure on the local force isosurface. In the `saddle' geometry the phenyl rings are more strongly tilted than in the `inverted' geometry, and the neighbouring macrocycle produces a shoulder in the isosurface that affects the plane fit.

The measured phenyl-ring angles for both 2H-TPP geometries are plotted in Figure \ref{planeFitResults}c, alongside the benzene and pyrrole results. This comparison shows the transition from idealised molecular fragments to a larger non-planar molecule. As previously discussed for benzene and pyrrole, the fitted force isosurface closely follows the imposed molecular tilt, allowing the input angle to be recovered quantitatively. For 2H-TPP, the same approach still distinguishes the `inverted' and `saddle' adsorption geometries, but the phenyl-ring angle in the `saddle' geometry is systematically underestimated. This demonstrates that force-isosurface images can retain clear conformational information in realistic non-planar molecules, while also showing that quantitative fitting must be interpreted with care when the local force field is influenced by neighbouring molecular structure.

\paragraph{Metal-centred porphyrins and phthalocyanines. }

The previous examples show that force isosurfaces can distinguish molecular adsorption geometries and recover the orientation of tilted molecular fragments. We next considered molecules where the key structural variation occurs near the centre of the molecule. This provides a different test of constant-force imaging, whether the force isosurface can retain contrast from chemically important central features while also revealing lower-lying parts of the molecular structure. 

\begin{figure}[h!]
\centering
\includegraphics[width=0.49\textwidth]{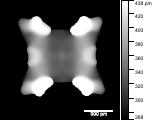}
\caption{Constant-force image of ZnTPP in the `saddle' adsorption geometry. The zinc atom occupies the centre of the porphyrin macrocycle and produces a raised cross-like feature extending towards the surrounding nitrogen atoms.\label{ZnTPP}}
\end{figure}

Porphyrins can host metal ions at the centre of the macrocycle in place of the two central hydrogen atoms, giving rise to a wide range of metalloporphyrins with distinct structural and electronic properties \cite{porphyrinReview}. Albrecht \textit{et al.} showed that the central region of adsorbed porphyrins can be distinguished by NC-AFM, with CuTPP producing a filled centre compared with the central depression observed for 2H-TPP \cite{JaschaRepp}. The simulated force isosurface of ZnTPP in the `saddle' adsorption geometry is shown in Figure \ref{ZnTPP}. The zinc atom produces a raised central feature, with contrast extending towards the four surrounding nitrogen atoms. This gives a cross-like structure in the centre of the molecule, consistent with the contrast expected for a metallated porphyrin.

\begin{figure}[h!]
\centering
\includegraphics[width=0.5\textwidth]{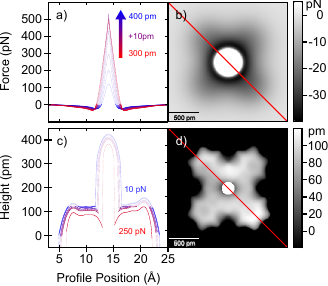}

\caption{Simulated constant-height and constant-force images of CO-FePc. In the constant-height image simulated at $300$~pm \textit{(b)}, the contrast is dominated by the protruding CO ligand and the FePc core is only weakly resolved. The corresponding force profile \textit{(a)} shows the larger force contribution above the CO ligand. In the constant-force simulation \textit{(d)}, the tip follows the selected force threshold and approaches the lower-lying parts of the FePc molecule, revealing submolecular contrast in the molecular core. The corresponding height profile is shown in \textit{(c)}.\label{CO-FePc_fig}}
\end{figure}

A related challenge is presented by carbon monoxide-functionalised iron phthalocyanine (CO-FePc). FePc is a relatively planar molecule and is therefore well suited to constant-height imaging. However, adsorption of carbon monoxide above the central Fe atom produces CO-FePc, where the protruding CO ligand obscures the underlying phthalocyanine core in constant-height images. Chen \textit{et al.} showed that progressively reducing the tip height can remove the CO ligand, allowing the FePc molecule to be imaged directly \cite{FePC-CO}. 	Using the atomic coordinates reported in that work, we simulated constant-height and constant-force images of CO-FePc, shown in Figure \ref{CO-FePc_fig}. The simulated constant-height image is consistent with the previously reported experimental and simulated images, with the contrast dominated by the protruding CO ligand. At this height, the phthalocyanine core lies sufficiently far from the tip that it contributes only weakly to the image. In contrast, the constant-force simulation allows the tip to approach the lower-lying molecular framework while maintaining the selected force threshold. This reveals submolecular contrast within the phthalocyanine core, including depressions in the isoindole sections.

Together, ZnTPP and CO-FePc show that force-isosurface contrast is sensitive to central molecular structure as well as overall molecular shape. In ZnTPP, the central metal atom produces a distinct raised feature in the force isosurface. In CO-FePc, a protruding axial ligand dominates constant-height contrast, but constant-force imaging recovers information from the lower-lying molecular core. These examples demonstrate how constant-force simulations can identify which structural features are expected to dominate the image and which lower-lying features may remain accessible.

\paragraph{Buckminster fullerene.}

Buckminster fullerene (C$_{60}$) provides a stringent test case for constant-force imaging because its strongly curved structure gives substantial height variation within a single molecule. C$_{60}$ has previously been imaged by constant-height NC-AFM \cite{jarvisc60, c60GrossBondOrder}, as well as by approaches designed to retain high-resolution contrast while following the molecular topography \cite{pawlak2012high, OscarCustance}. In these studies, force information could be reconstructed from $\Delta f$ data using methods such as the Sader-Jarvis formalism, or acquired along a topographic path defined by an initial AFM measurement \cite{pawlak2012high, SaderJarvis, OscarCustance}.

\begin{figure}[h!]
\centering
\includegraphics[width=0.5\textwidth]{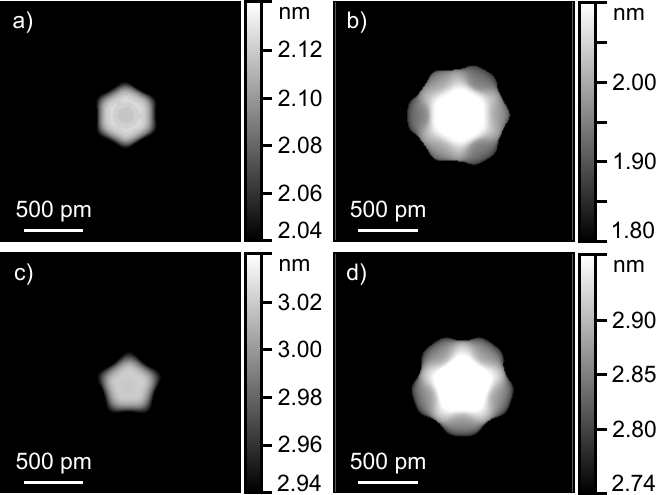}

\caption{Simulated constant-force images of C$_{60}$ in hexagon-up and pentagon-up orientations. With the colour scale focused on the uppermost hexagonal face \textit{(a)}, the top face is clearly resolved. Adjusting the colour scale to lower heights reveals the six neighbouring faces, with distinct contrast from adjacent pentagonal and hexagonal faces \textit{(b)}. In the pentagon-up orientation, the uppermost pentagonal face is resolved \textit{(c)}, together with the five neighbouring hexagonal faces \textit{(d)}.\label{bucky_fig}}
\end{figure}

The simulated constant-force images of C$_{60}$ are shown in Figure \ref{bucky_fig}. For both hexagon-up and pentagon-up orientations, the uppermost molecular face is clearly resolved. The adjacent faces are also visible in the force isosurface, demonstrating that constant-force contrast can retain structural information over the curved molecular surface. In the hexagon-up orientation, the neighbouring pentagonal and hexagonal faces produce distinct contrast, allowing the molecular orientation to be identified beyond the uppermost face alone.

\paragraph{Conclusion.}

We have used probe-particle-model simulations to predict the constant-force contours expected above a range of molecular systems. These simulations show that force isosurfaces can retain structural information that is lost in conventional constant-height measurements when molecules have significant topographic variation. In tilted benzene and pyrrole, the molecular structure remains visible across the range of angles studied, allowing molecular orientation to be recovered quantitatively. For larger non-planar molecules, the same approach identifies the characteristic force-isosurface contrast associated with different adsorption geometries, lower-lying molecular features and strongly curved molecular surfaces.

The results also clarify how constant-force molecular images should be interpreted. A constant-force image is not a direct topographic map of the molecular skeleton, but a surface determined by the full tip–sample interaction field. For simple molecular units this force isosurface can provide quantitative structural information, while in more complex molecules neighbouring features can influence the measured contour. Simulations are therefore important for assessing which structural features are likely to be accessible, and for distinguishing geometric information from contrast arising from the surrounding force field.

These simulations provide target contrasts for constant-force isosurfaces that could be extracted from three-dimensional force-mapping experiments, helping to assess when such experimentally demanding measurements are likely to provide useful structural information. They also provide a practical method for interpreting constant-force molecular AFM images acquired using force-feedback approaches such as off-resonance or PeakForce-style measurements. In this way, force-isosurface simulation can support the interpretation and design of experiments aimed at resolving three-dimensional structure in non-planar molecular systems.

%
%

\paragraph{Acknowledgements.}

The authors would like to thank Prokop Hapala for helpful discussion about the Probe Particle Model. S.P.J. thanks the Engineering and Physical Sciences Research Council (EPSRC) and the Royal Society, respectively, for grants EP/X026876/1  and PI70026. E.J.D., S.P.J., and R.J.Y. gratefully acknowledge funding from the Graphene NOWNANO CDT, EP/L01548X/1.

\bibliographystyle{vancouver}
\bibliography{references_corrected}

\end{document}